\documentclass[titlepage,aps,reprint,showpacs,pra,amsmath,amssymb,floats,floatfix,superscriptaddress,]{revtex4-2}

\newcommand{\vbf}[1]{\mathbf{#1}}

\newcommand{\vct}[1]{{\bf #1}}
\newcommand{\eq}[1]{Eq.~(\ref{eq:#1})}
\newcommand{\fig}[1]{Fig.~\ref{Fig:#1}}

\DeclareMathAlphabet{\mathpzc}{OT1}{pzc}{m}{it}

\usepackage{inputenc}
\usepackage{ulem}
\usepackage[T1]{fontenc}
\usepackage{cases}
\usepackage{graphicx}
\usepackage{dcolumn}
\usepackage{bm}
\usepackage[usenames,dvipsnames]{xcolor}
\usepackage{mathrsfs}
\usepackage{scalefnt}
\usepackage{hyperref}
\usepackage[caption=false]{subfig}

\hypersetup{
    colorlinks=true,
    linkcolor=blue,
    filecolor=cyan,      
    urlcolor=purple,
    citecolor=orange
}

\definecolor{cgreen}{RGB}{0,64,00}
\definecolor{cblue}{RGB}{86,180,233}
\definecolor{cred}{RGB}{229,30,16}
\definecolor{cpurple}{RGB}{148,0,211}
\definecolor{cyellow}{RGB}{240,228,66}
\definecolor{corange}{RGB}{230,159,0}
\let\xxxhat\hat

\renewcommand{\hat}[1]{{\boldsymbol {\xxxhat {#1}} }}

\begin{document}

\title{Finite-range effects in the unitary Fermi polaron}

\author{Renato Pessoa}
\affiliation{Instituto de F\'{\i}sica,
Universidade Federal de Goi\'as - UFG,
74001-970 Goi\^ania, GO, Brazil.} 

\author{S. A. Vitiello}
\affiliation{Instituto de F\'{\i}sica Gleb Wataghin,
Universidade Estadual de Campinas - UNICAMP,
13083-970 Campinas, SP, Brazil.}

\author{L. A. Pe\~na Ardila}
\affiliation{Institut f\"ur Theoretische Physik, Leibniz Universit\"at, 30167 Hannover, Germany.}

\date{\today}
\begin{abstract}
Quantum Monte Carlo techniques are employed to study the properties of
polarons in an ultracold Fermi gas, at $T= 0,$ and in the unitary regime
using both a zero-range model and a square-well potential. For
a fixed density, the potential range is varied and results are
extrapolated and compared against a zero-range model. A discussion regarding the choice of an
interacting potential with a finite range is presented. We compute the polaron effective mass, the polaron binding energy, and the effective coupling between them. The latter is
obtained using the Landau-Pomeranchuk's weakly interacting
quasiparticle model. The contact parameter is estimated by fitting the pair distribution function of atoms in different spin states.

\end{abstract}


\maketitle

\section{Introduction}

One of the most paradigmatic and appealing problems in physics is related to impurities interacting with a strongly correlated many-body environment, not only becauseof  its complexity, but also because it serves as a testbed of fundamental ideas such as the Landau-Fermi theory~\cite{LandauFermiLiquidTheory}. The concept of the polaron as a quasiparticle, first devised by Landau and Pekar~\cite{Landau48}, offers an alternative to explain several properties of materials in terms of collective excitations. Impurity particles interacting with a medium
can form polarons which behave like a single particle with renormalized properties such as
energy and mass~\cite{Frohlich54,fey54,PolaronsandBipolarons}. Transport properties in materials are
understood in terms of quasiparticles, for instance, the colossal
magnetoresistance
~\cite{Mannella05}, transport in organic materials~\cite{Gershenson78,Watanabe14}, or the Kondo effect due to pinned magnetic impurities~\cite{Kondo64}.

 The realization and control of
quasiparticles is attainable with current state-of-the-art experiments
in ultracold quantum gases. Depending on the statistics of the host bath,
polarons can be either Bose or Fermi polarons. The latter can be formed by a
spin-down impurity immersed in a polarized Fermi sea made of
 $N_{\uparrow}$ spin-up
particles. This is the simplest system for studying strongly
correlated imbalanced mixtures of fermions. The Fermi polaron problem has
attracted much interest from the theoretical and experimental point
of view~\cite{mas14,par06,zwi06,sch07,sch09,bug13,bra15,yan19,nes20,cet16,kos12,sca17,cui20}, in particular for its direct analogy with other systems
in the context of solid-state~\cite{Sidler17,TanLi2020} and nuclear~\cite{Vidana21} physics. 

Active research in polaronic physics in ultracold quantum gases is possible due
to the high versatility in controlling the strength of interactions via
Feshbach's resonances~\cite{cheng10}. In the low-energy regime, the natural length scale is the $s-$wave scattering length between particles of opposite spins and the Pauli principle exclusion restricts the direct interaction between particles in the Fermi sea. In the ultra dilute regime, the range of the interatomic potential of a strongly interacting Fermi gas is of the order of an effective radius (as defined by the low-energy expansion of the \textit{s}-wave scattering amplitude) and is much smaller than the interparticle distance $k_{F}\sim n^{-1/3}$. As a result, the only remaining length scale is the density, and the low-energy scattering of atoms in different spin states is unable to probe details of the interatomic interaction. In this regime, the properties of the system are said to be universal. Likewise, finite-range effects on the Bose polaron problem have  also been  investigated recently~\cite{Massignan2021,Levinsen2021,Enss2021}.

A physical situation where properties of interest do not depend on the details of the interatomic potential allows for the use of simple effective potentials. An example is given by the square-well potentials in numerous theoretical studies~\cite{ast04,ast05,lob06,lob06ns,pil08,ast12,ard15,sch17,ard16}. However, it is still necessary to extrapolate $R_0 \rightarrow 0$, or to choose an $R_0$ small enough with respect to the interparticle distance, to enable the proper estimation of the properties of the system. Nevertheless, the simplest approach is to use a model where the zero range is strictly enforced. Within the Wigner-Bethe-Peierls model, a zero-range model can be used where a contact potential is replaced by a condition in the many-body wave function~\cite{gio08}.

In this work, we consider the strongly interacting unitary Fermi gas where the scattering length
$|a| \rightarrow \infty$ and we consider the extremely imbalanced case, namely, the Fermi polaron, also called the $N+1$ system. By using fixed-node diffusion Monte Carlo, we study the ground-state properties of both the $N+1$ and $N+M$ problem, with $M$ the number of impurities, as a function of the range of the potential. A stringent evaluation of how an attractive short-range square-well potential might impact the results is made by considering the zero-range model. Results for the binding energy, effective mass, and the coupling of polarons in a unitary ultracold Fermi gas are obtained. The contact parameter of the gas is found through the pair distribution
function of unlike spin particles. The zero-range limit of this interaction is compared with the zero-range model.

\section{System and Method}

The excitation spectrum of a "slow" Fermi polaron  $\mathbf{p}\ll \hbar \mathbf{k}_{F}$ (with $\hbar \mathbf{k}_F$ the Fermi momentum) that behaves as a free quasiparticle of effective mass $m^*$ is given by
    \begin{equation}
        \Delta E =\frac{\mathbf{p}^2}{2m^*}-\frac{3}{5}E_{F\uparrow}A,
        \label{eq:Emodel}
    \end{equation}
where
$3E_{F\uparrow}/5$
is the full polarized free Fermi gas energy $E_{F\uparrow}=\hbar^2k_F^2/ 2m$ of atoms with mass $m$, $A$ is a universal parameter when the interatomic potential range tends to zero, and $k_F=(6\pi^2n_\uparrow)^{1/3}$, where $n_\uparrow$ is the bath density. At the unitary limit, the polaron binding energy  $-3E_{F\uparrow}A/5$ is proportional to the bath energy because the atomic density is the only relevant length scale in the system. 

Beyond the single impurity regime, a partial polarized normal gas is characterized by a concentration $x=N_\downarrow/N_\uparrow$ of down spins $N_\downarrow$ with respect to up spins $N_\uparrow$, and the system ground-state energy is given by the Landau-Pomeranchuk model~\cite{pil08,mora10},
\begin{equation}
    \frac{E}{N_{\uparrow}}=\frac{3}{5}E_{F\uparrow}\left( 1 - Ax +
    \frac{m}{m^*}x^{5/3}+Fx^2\right),
    \label{eq:eos}
\end{equation}
where $F$ accounts for interactions between polarons. The equation of state with $F=0$ was used to describe the system in small concentrations of impurities, in  good agreement with the polaron properties estimated with $F\neq 0$~\cite{lob06ns,pil08}. In addition, the tail of the momentum distribution of a single spin component interacting Fermi gas $n_\sigma(\vct k)$ is
related to the universal contact parameter \cite{tan08a}  $C$ via
\begin{align}
C = \lim_{\vct k \rightarrow \infty} k^4 n_\sigma(\vct k).
\label{eq:tanc}
\end{align}
This parameter is a measure of short-range correlations among atoms in different
spin states for a given concentration~\cite{jen20,dus12,tan08b,tan08a},
\begin{equation}
g(x,k_F r\rightarrow 0)
= \frac{C}{16\pi^2 n_{\uparrow}n_{\downarrow}}  \frac{1}{r^2}.
    \label{eq:gtan}
\end{equation}
 In a homogeneous case, the density of a single impurity component and the bath scales as $n_\downarrow=1/\Omega$ (with $\Omega$ the volume) and 
$n_\uparrow = k_F^3/6\pi^2,$ respectively, and the contact reduces to 
\begin{align}
g(k_F r\rightarrow 0)
    = \frac{3}{8} \frac{\mathcal C}{k_F}\frac{1}{(k_F r)^2},
    \label{eq:gpol}
\end{align}
where the dimensionless contact per unit volume,
${\cal C}/N_\uparrow k_F$, is
related to $C$ through
${\cal C}/N_\uparrow k_F = 6\pi^2 C / k_F^4$.

\subsection{Trial wave functions}
The structure of the wave function to treat a normal ultracold Fermi gas follows the general form given by
\begin{equation}
    \Psi(\mathbf{R}) =
    \prod_{i}^{N_\uparrow}\prod_{i'}^{N_\downarrow} f(r_{ii'})
    \Phi
    \varphi,
    \label{eq:psi}
\end{equation}
where $\mathbf{R}$ is the configuration of the atoms, $\mathbf{R}=\{\vbf{r}_{1},\vbf{r}_{2},\ldots,\vbf{r}_{N_\uparrow},\vbf{r}_{1'},\vbf{r}_{2'} ,\ldots, \vbf{r}_{N_\downarrow}\}$, the unprimed (primed) index depicts the up spins (down spins) of
$N_\uparrow$ $(N_\downarrow)$ particles, $f(r_{ii'})$ is a model-dependent Jastrow factor that depends on the relative distance $r_{ii'} = \vert \vbf{r}_i - \vbf{r}_{i'} \vert$ that correlates the minority with the
majority up spin atoms, $\Phi$ is a Slater determinant of plane-wave
orbitals describing the up spin atoms, and $\varphi$ describes the
minority atoms wavefunction. The latter can be either a single plane wave for a single impurity or a
Slater determinant of plane waves for different concentrations $x$ of down-spin impurities. In both $\Phi$ and $\varphi$, atoms are described
by plane waves with wavevectors given by
$k=\frac{2\pi}{L}(n_x^2+n_y^2+n_z^2)^{1/2}$, where $n_x,$ $n_y$, and
$n_z$ are integer numbers and $L=\Omega^{1/3}$ is the side of the
simulation cell. The specific functional form of the wave functions will be discussed in the following.

\subsection{Short range square-well potential}

From the theoretical point of view, the unitary
regime allows us to change
the real interatomic potential
by an effective potential which
is simple
and captures the most
important features of low-energy scattering.
For finite-range potentials, then the limit of zero range can be
taken. The Hamiltonian of the system with
$N=N_\uparrow+N_\downarrow$
atoms can be written as
\begin{equation}
H = -\frac{\hbar^2}{2m}\left[ \sum_{i}^{N_\uparrow}\nabla_{i}^2
    + \sum_{i'}^{N_\downarrow}\nabla_{i'}^2 \right]
    + \sum_{i,i'} V(r_{ii'}),
    \label{eq:H}
\end{equation}
where $V$ is the interacting potential between unlike-spin pairs that depends on the relative distance
$r_{ii'}$.
For the single polaron, of course, $N_\downarrow = 1$ and we do not have
the primed sums but only a single term that corresponds to the impurity,
immersed in the gas at a chosen momentum state.

\begin{figure*}[htp]
\centering
     \subfloat{\label{Fig:deltaE}\includegraphics[scale=0.5]{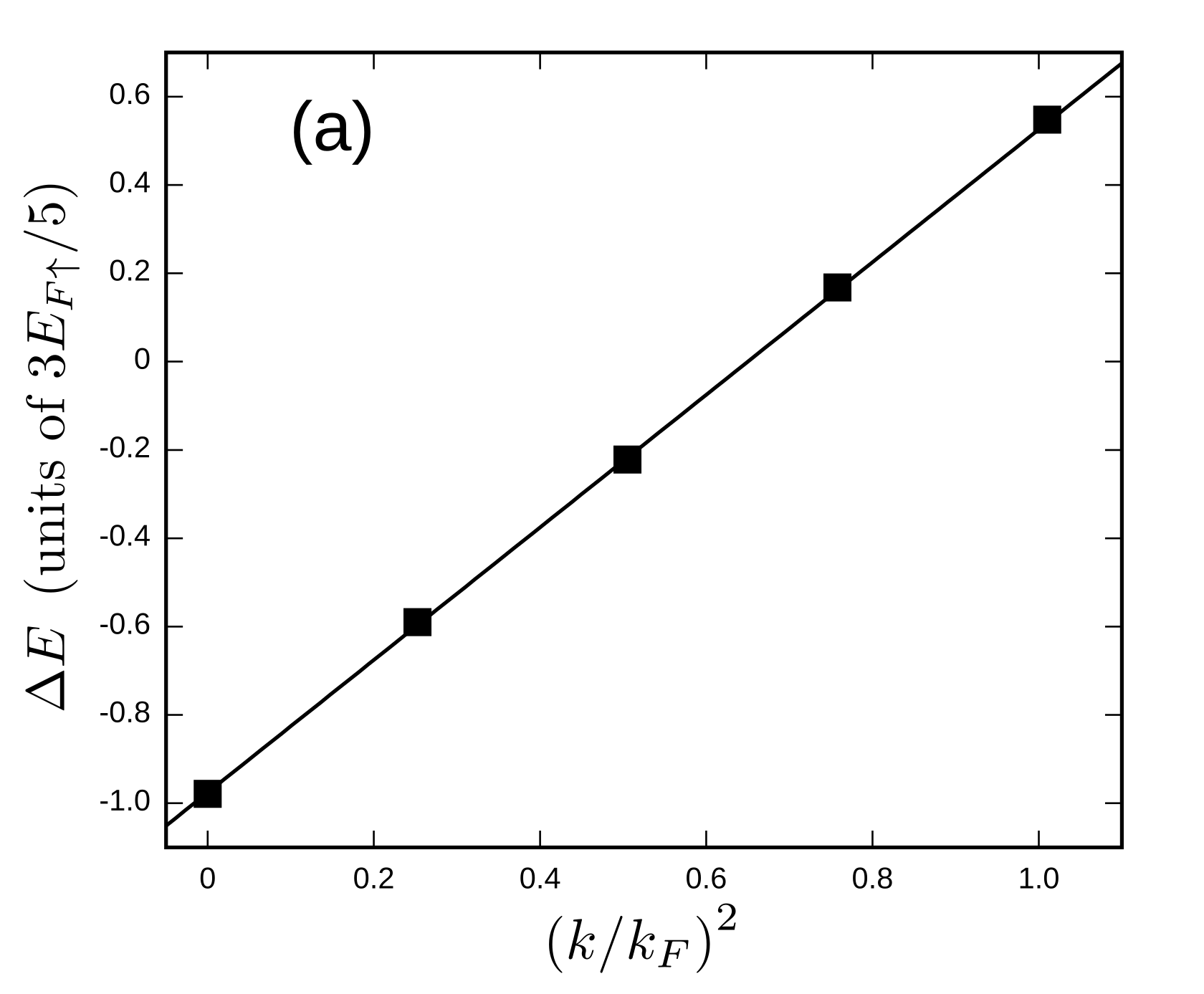}}
     \subfloat{\label{Fig:deltaEsw}\includegraphics[scale=0.5]{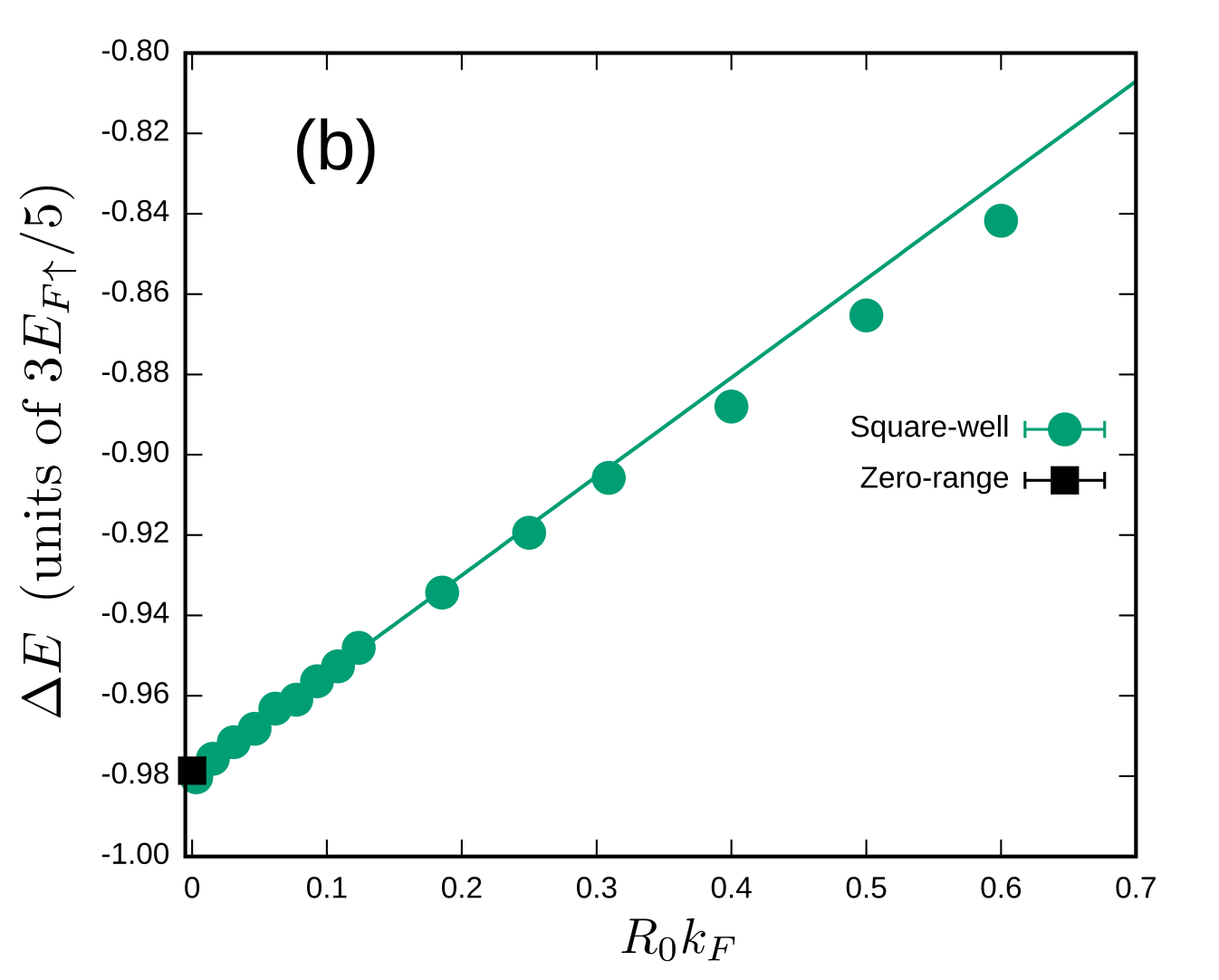}}
\caption{(a) The polaron excitation spectrum for the zero-range model. The solid line is a fit performed in the range $0\le (k/k_F)^2 \le 0.6$. (b) The difference of energies between the system with the impurity and the fully polarized system as a function of the square-well potential range $R_0k_F$. The line is a fit to the estimated values considering the range $0 \le R_0 k_F \le 0.2$. 
}
\label{Fig:delE}
\end{figure*}

At low energy, the details of a short-range interaction are not relevant, allowing for the use of different effective potentials. A customary potential employed in many quantum Monte Carlo
simulations~\cite{ast04,ast05,lob06,lob06ns,pil08,ast12,ard15,sch17,ard16} is the spherical square-well potential,
\begin{eqnarray}
 V(r) &=& \left \{ \begin{array}{l r}
     -V_0  ,\;\; \;\;  &  r < R_0, \\
     0, &    r \ge R_0 .
 \end{array}\right.
 \label{eq:V_sw}
\end{eqnarray}
where $R_0$ is the potential range and $V_0$ its strength. The unitary
regime is obtained by imposing
$V_0=(\pi/2R_0)^2$.
This is the condition for the appearance of a first bound state in the
potential well, causing the scattering length to go to infinite.

Since the atomic scattering is not able to probe internal structures in the investigation of ultracold atomic gases, the parameter $A$ of \eq{eos} exhibits a universal linear dependence in the
product of the effective range multiplied by the Fermi momentum
\cite{cas11,wer12} for small effective ranges $R_0$,
\begin{align}
    A(R_0k_F) = 
\mathcal{A} + \eta R_0 k_F + {\cal O}(R_0 k_F)^2,
    \label{eq:Aur}
\end{align}
where $\mathcal{A}$ and $\eta$ are parameters to be estimated.

The Jastrow factor in the wave function of \eq{psi} for relative distances smaller than the range $R_0$ is a zero-energy solution for the two-body Schr\"odinger equation in the square-well potential, whereas, for distances greater than $R_0$, the function used obeys the boundary condition for short distances in the limit of the range equal to zero, i.e, $f(r)\propto 1/r$ and
         \begin{subnumcases}{f(r) =}
          \ensuremath{\mathpzc{A}\frac{\sin(k_0 r)}{r},}   &  $r < R_0,$
             \label{eq:near} \\
          \ensuremath{\mathpzc{B}\frac{\cosh(\lambda r)}{r},} & $R_0 \le r \le D$.
             \label{eq:far}
             \end{subnumcases}
The multiplicative constant $\mathpzc{A}$ is chosen such that $f(r)$
is continuous at $r=R_0$ and $k_0 = \sqrt{mV_0}/\hbar$. The parameter $D \leqslant L/2$ is the healing
distance, which is optimized to obtain the lowest variational energy.
Coefficient $\mathpzc{B}$ and parameter $\lambda$ are chosen in order that $f(r=D)=1$ and $f'(r=D)=0$.

There are other choices in the literature for the Jastrow factor for Fermi gases interacting with the square-well potential~\cite{ast04,ast05,sch17}. However, the chosen form introduces, in a simple way, the most important features of the wave function for this potential. 

\subsection{Zero-range model}

The zero-range limit can be taken by only considering the low-energy
scattering length. This is done by replacing the actual atomic interatomic potential with the Wigner-Bethe-Peierls contact condition in the $N$-body wave-function. Within this approach, the Jastrow factor of
\eq{psi} has the same functional form in the whole range of up-down
spin separations,
\begin{equation}
f(r)= \mathcal{N} \frac{\cosh(\lambda r)}{r}.
\end{equation}
Here, the function $f(r)$ at $r=D$ is normalized by the constant $\mathcal{N}$. The functional form  adopted here for $f(r)$ is similar to the one in \eq{far}, but does not involve a potential range. Moreover, since we are
interested in the Fermi gas with infinite scattering length, the contact
interaction is taken into account by the
Wigner-Bethe-Peierls boundary condition by imposing $\Psi(r_{ij'}\rightarrow 0)=\frac{1}{r_{ij'}}$ for particles of different spins. The Hamiltonian for wavefunction $\Psi$ is identical to
the one for the ideal gas $H = -\frac{\hbar^2}{2m}\sum_{i}^{N}\nabla_{i}^2$, where the sum considers both spin states~\cite{pes19}.

\section{Results and Discussion}

The polaron binding energy can be straightforwardly computed by
considering the zero-range model. From the total energy of the impurity-bath system computed with the down-spin particle in the zero momentum state, we subtract the polarized
Fermi gas energy of $N_\uparrow$ particles. Thereby, that difference provides us  with an estimated value of $-3E_{F\uparrow}A/5$ and this allows an estimation of the universal parameter
$A=0.97861(57)$. An alternative way to estimate this quantity is done by using the excitation spectrum $\Delta E$ in \eq{Emodel} obtained by considering the impurity in different momenta states $\mathbf{p}=\hbar \mathbf{k}$. The results as a function of $(k/k_F)^2$ are fitted to a straight line and its value at $\mathbf{k}=\mathbf{0}$ is used to estimate $A$; see \fig{deltaE}. The obtained value gives $A=0.9758(70)$ which is in excellent agreement with its direct calculation in terms of the impurity chemical potential. The fitted line also allows the estimation of the effective mass $m^*/m = 1.1101(18)$. The excitation spectrum of the
quasiparticle model in \eq{Emodel} is used to fit in the range $0 \le \left( k/k_F\right)^2 \le 0.6$. Note that the excitation spectrum as a function of $\left( k/k_F\right)^2$ does not considerably deviate  from a linear behavior in the whole range depicted in \fig{deltaE}.

\begin{figure}[htp]
 \centering
     \subfloat{\label{Fig:pol_m}\includegraphics[scale=0.5]{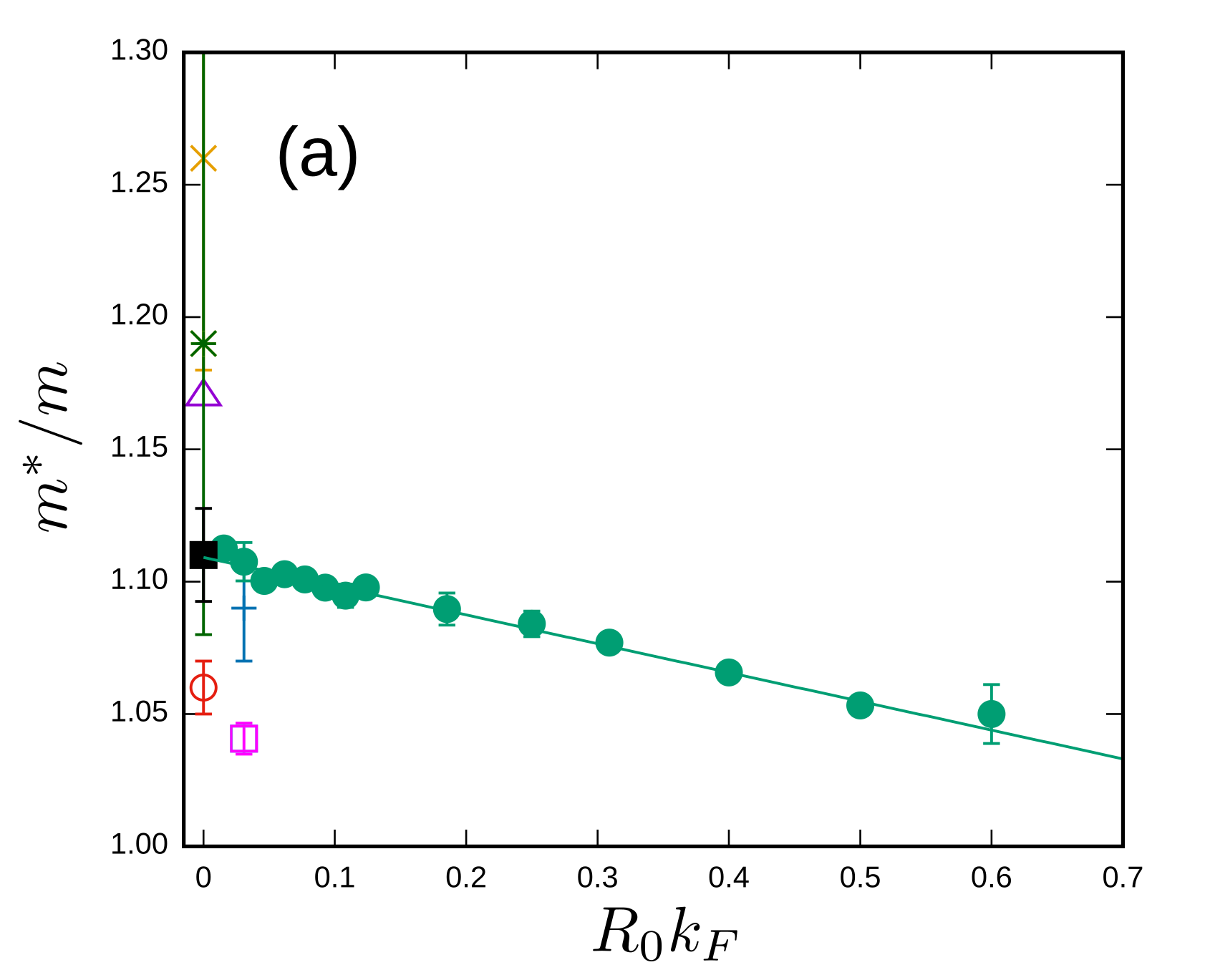}}\\
     \subfloat{\label{Fig:pol_A}\includegraphics[scale=0.5]{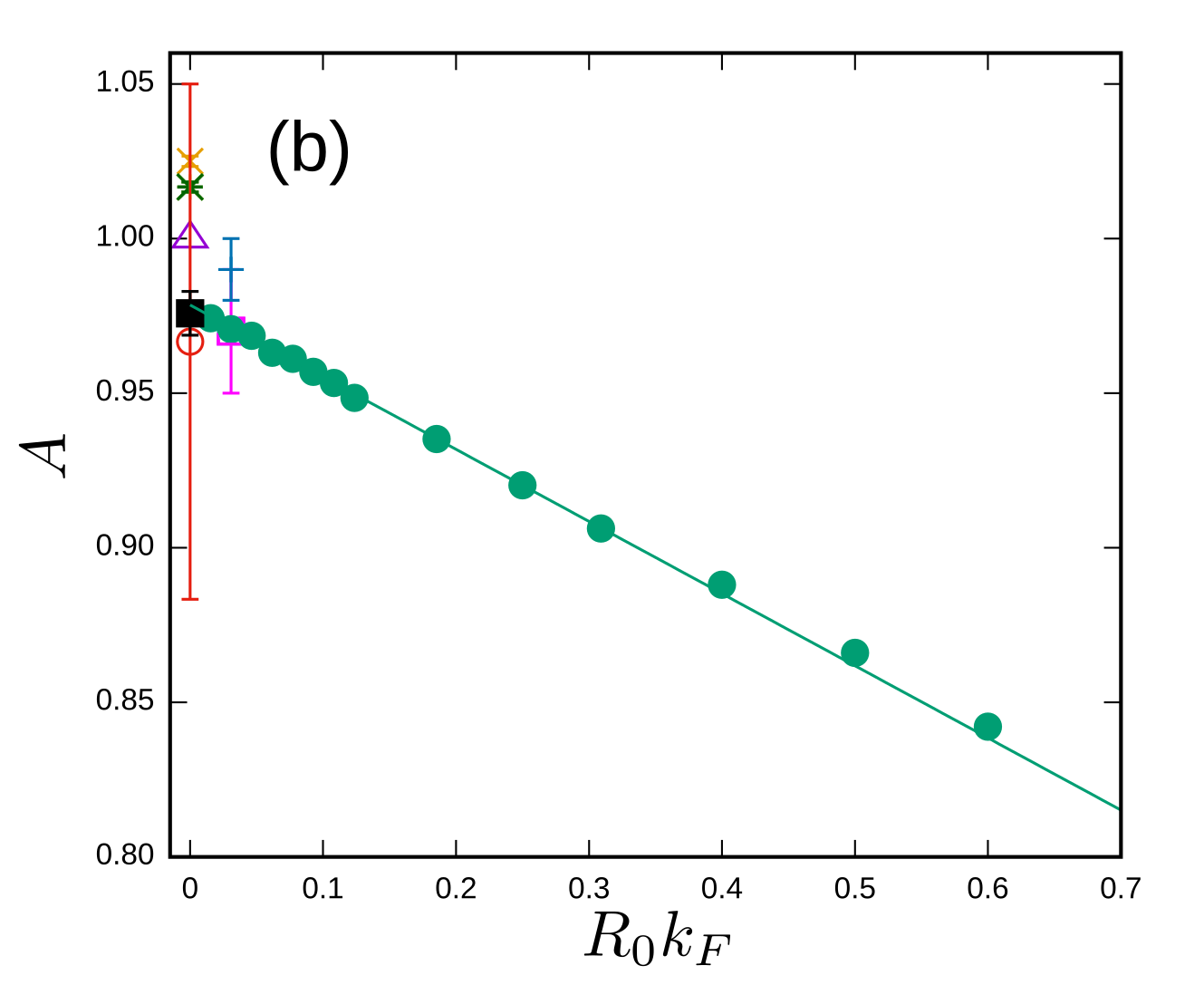}}
    \caption{
        (a) Effective mass of polaron as a function of the potential range.
        (b) Values of parameter $A$ extrapolated to $\mathbf{k}=\mathbf{0}$. The zero-range model results are included for the purposes of comparison with the square-well potential  and are depicted by black filled squares ($\blacksquare$). Results from the literature given by Lobo \textit{et al.}~\cite{lob06ns} and Pilati \textit{et al.}~\cite{pil08} are shown by empty squares ($\color{magenta} \square$) and the plus ($\color{blue}+$), respectively. Empty triangles ($\color{purple}\triangle$) represent the analytical results obtained with the Chevy ansatz~\cite{che06,com07}. Diagrammatic results of Vlietinck \textit{et al.}~\cite{vli13} are represented by times ($\color{orange}\times$) and experimental results are presented by empty circles ($\color{red}\circ$)~\cite{shi08} and asterisk ($\color{cgreen}*$)~\cite{nas10}.
     \label{Fig:pol_m_A} 
}
 \end{figure}

For simulations using the square-well potential, the total energy of a system with a single impurity as a function of the potential range subtracted from the full polarized free Fermi gas energy is shown in
\fig{deltaEsw}. A linear fit to the results and the extrapolation
$R_0 \rightarrow 0$
gives the estimation of $0.97916(55)$ for $\mathcal{A}$.
Additionally, for each of the
ranges considered, simulations with the impurity in different momenta
states were also performed. Similarly as before, we can estimate the parameter $A$ and the effective mass
$m^*/m$ as a function of the potential range  (see \fig{pol_m_A}). Linear fits that were extrapolated to
$R_0 \rightarrow 0$
gives $A=0.9785(11)$ and $m^*/m = 1.1085(15)$, values that are in excellent agreement with the results for the zero-range model,
also displayed in \fig{pol_m_A}. The trend of
these estimated quantities using the effective potential converges to the values obtained with the zero-range model, as expected.

Results for the parameter $A$ displayed in \fig{pol_A} scale linearly to its value at $R_0k_F \rightarrow 0$.
This is strong evidence that
the universal linear relation of parameter $A$ as a function
of the range of an effective potential in \eq{Aur}
is verified. Extrapolated results for $A$
in the range $0 \le R_0 k_F \le 0.2$ are
fitted to \eq{Aur} up to first order in $R_0 k_F$ and extrapolating to $R_0 \rightarrow 0$. This gives the result $\mathcal{A}=0.97916(55)$, which is in excellent agreement with the value obtained from the zero-range model. These results show that the linear behavior between $A$ and $R_0 k_F$ is indeed fulfilled and therefore there is always a dependence on the range even if $R_{0}k_F\ll1.$\\

The fitted curve slope $\eta=-0.233(14)$ in \eq{Aur} [see, also, \fig{pol_A}] for the $N+1$ system differs considerably from the value obtained for  an unpolarized system determined with diffusion Monte Carlo (DMC) by Schonenberg and Conduit
 \cite{sch17}, $\eta=0.087(1)$ or by Carlson \textsl{et al.} \cite{car11}, $\eta=0.11(3)$, using auxiliary field Monte Carlo. These results show that in the calculations, here the potential range plays a less significant role for the unpolarized system than for the polaron problem. Both experimental and theoretical results from the literature are also presented in \fig{pol_m_A}. In general the agreement is quantitatively good. The closest value to our result of $m^*/m$ obtained using DMC in the literature~\cite{pil08,lob06ns} is reported in the more recent one.

\begin{figure}[htp]
\centering
\includegraphics[scale=0.5]{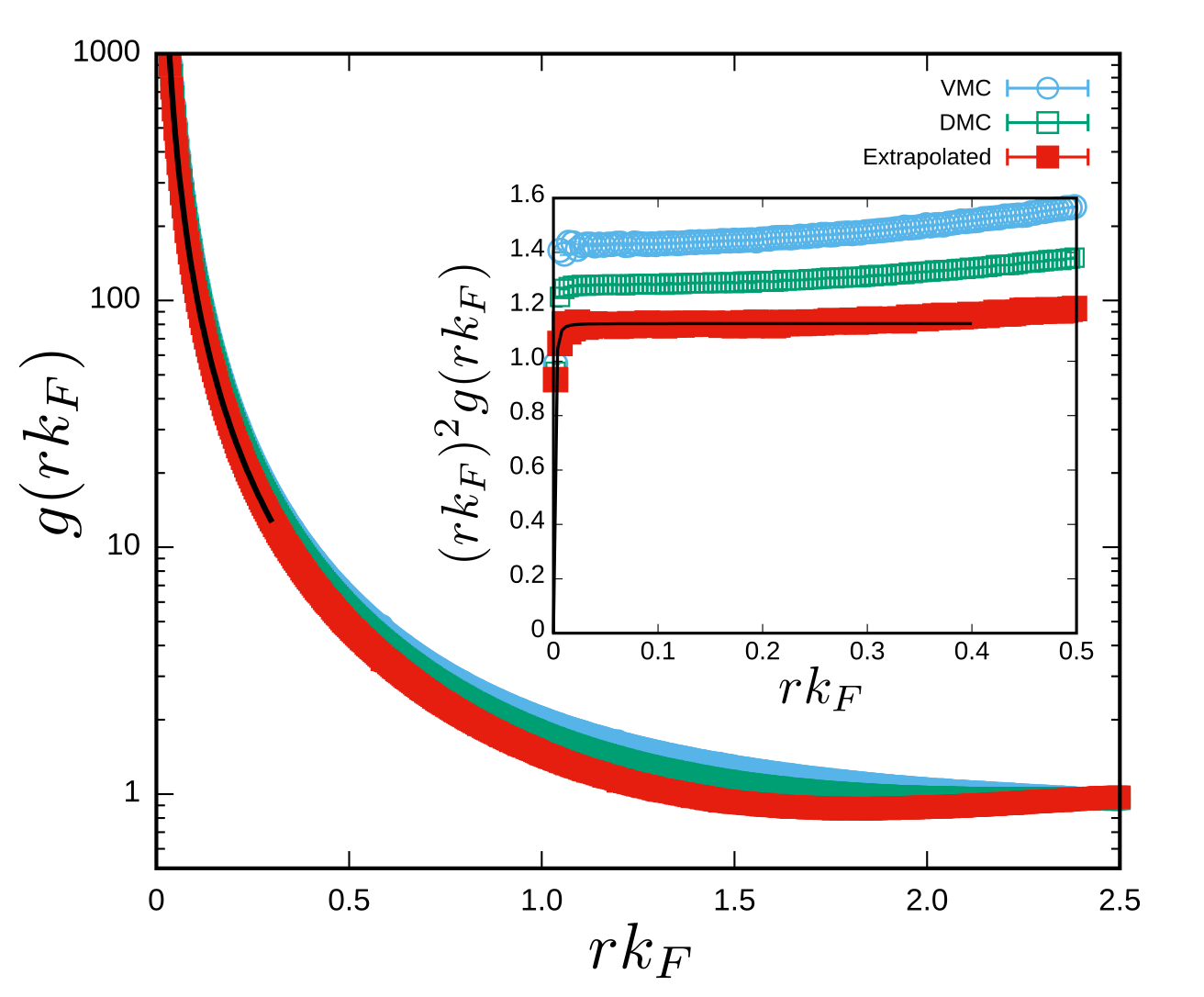}
\caption{Pair distribution function of unlike spins in the impurity-bath system as a function of the distance. The impurity is chosen initially at rest ($k/k_F=0$). Both variational Monte Carlo VMC and DMC results are extrapolated as customary for non pure estimators. The solid line is the best fit of the function $g(rk_F)= a/(rk_F)^2$ in the range $0 \leqslant rk_F \leqslant 0.4$. The inset shows the function $g(rk_F)(rk_F)^2$.}
\label{Fig:gr}
\end{figure}

The pair distribution function between a down-spin atom and the
polarized up-spin Fermi gas as a function of the relative distance
estimated using the zero-range model is shown in \fig{gr}. The
normalization is such that $g(rk_F \rightarrow \infty)\rightarrow 1$
and the extrapolated results, $g(rk_F)=2g_{\textrm{DMC}}-g_{\textrm{VMC}}$, are represented by the red squares. The contact parameter per unit volume is estimated by fitting
$(rk_F)^2g(rk_F)$ to a straight line in the range $0 \leqslant rk_F
\leqslant 0.4$ ~\cite{gan11,ard15,pes15pra,pes19}, as shown in the  of \fig{gr}. This figure also shows
a fit of $g(rk_F)$ to $a/r^2$ and the fits follow the expected behavior
$g(rk_F)\propto \frac{1}{(rk_F)^2}$ shown by \eq{gpol}.

\begin{figure}[htp]
\centering
\includegraphics[scale=0.5]{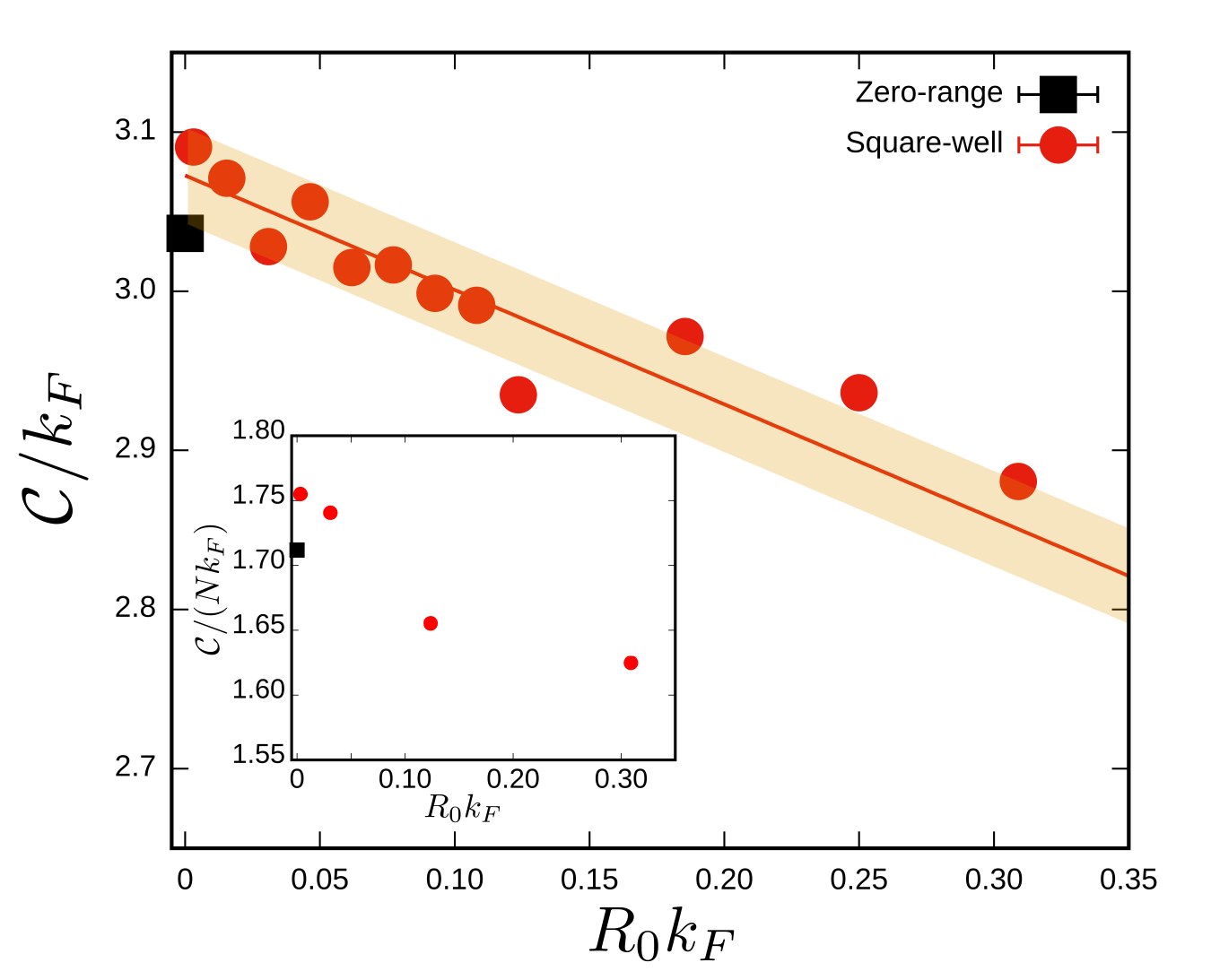}\caption{Contact parameter per unit of volume as a function of the potential range for one down-spin impurity in the bath of $N_{\uparrow}$ particles. The black square stands for the zero-range model result. The yellow region shows the dispersion of the square-well results depicted by red dots with respect to the fitted results. The inset shows results for the unpolarized Fermi gas in the normal state.}

\label{Fig:cont}
\end{figure}

From the pair distribution function, the estimated result using the zero-range model, $\mathcal{C}/ k_F=3.0363(21)$, is
depicted in \fig{cont} by a black square. In the same figure,  the
contact parameter per unit volume as a function of the range
of the square-well potential is also presented. The solid line is a fit of $\mathcal{C}/k_F$ to the function $\mathcal{C}/k_F (R_0k_F) = \mathcal{C}_0 + \mathcal{C}_1(R_0k_F)$. The fitted parameters are $\mathcal{C}_0=3.0727(92)$ and $\mathcal{C}_1=-0.719(99)$. The result $\mathcal{C}/k_F = 4.74$ from Ref.~\cite{lev17}
affected by a variational bias is
larger than ours.
Regarding experimental values,
a recent result obtained via rf spectroscopy indicates a contact
parameter of $\mathcal{C}/k_F = 4.0 \pm 0.5$ for a highly spin-unbalanced Fermi gas~\cite{yan19}. To verify the importance of the size effects, particularly in the
estimation of $\mathcal C$, a simulation with $N_{\uparrow} = 57$ atoms
in the bath was done.
We obtained $\mathcal{C}/ k_F=3.0294(16)$ giving an indication that
results for both $N_\uparrow = 33$ and $N_\uparrow = 57$ are
within the thermodynamics limit.

For completeness, the contact parameter per unit volume as a function of the potential range for the
unpolarized gas with $N_{\uparrow}=N_{\downarrow}=33$ particles is shown
in the inset, which also displays the estimated value using the
zero-range model. This value, $\mathcal{C}/Nk_F=1.7118(10)$, obtained for the Fermi gas in the normal state is significantly lower than the
one obtained in the superfluid state~\cite{gan11,pes19}. The same trend
is observed in the literature for both experimental and theoretical
results~\cite{jen20}, indicating, thus, an abrupt drop in the value of the contact parameter for the unpolarized system when the Fermi gas goes from the
superfluid to the normal phase.

\begin{figure}[htp]
\centering
\includegraphics[scale=0.5]{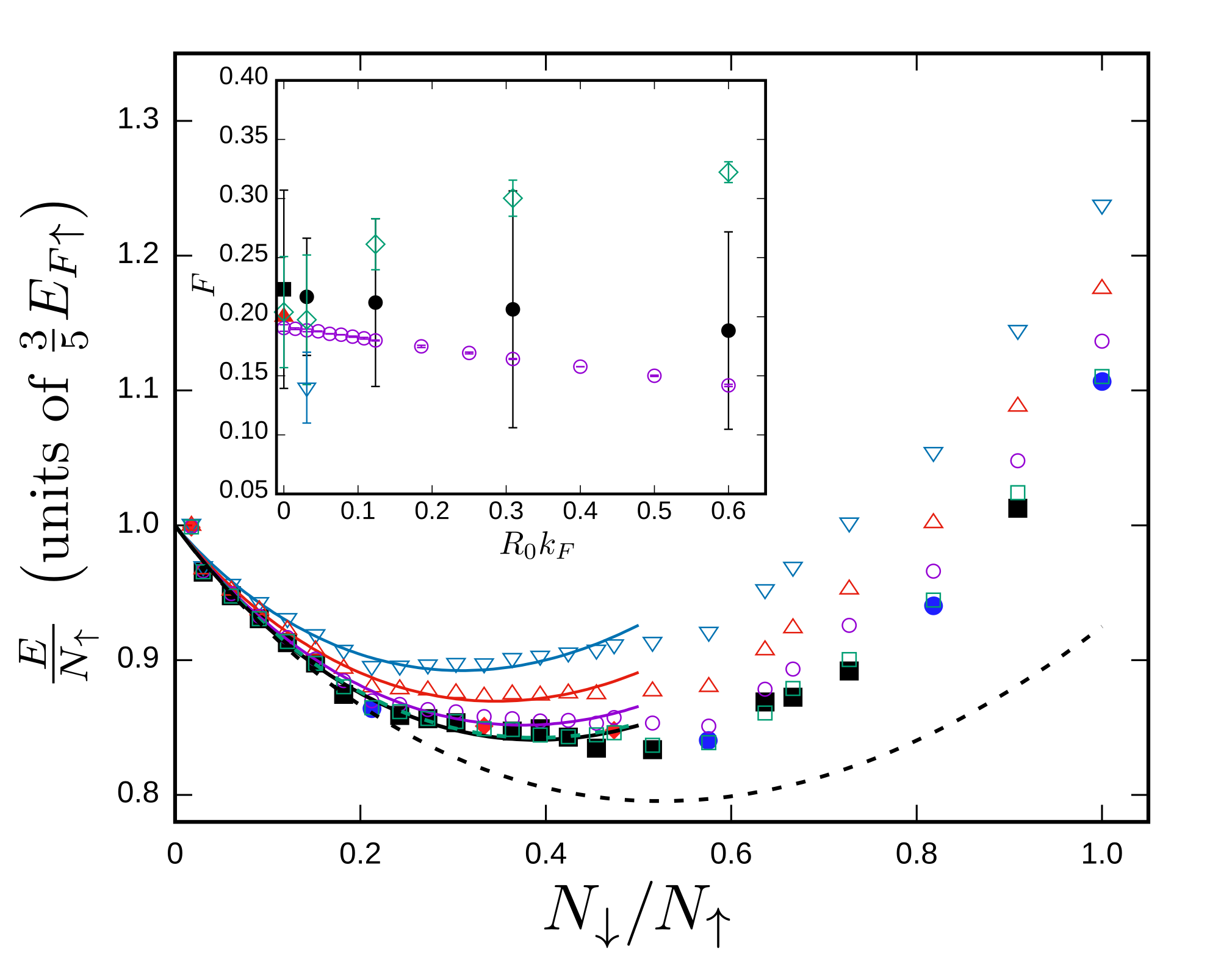}
\caption{Energy as a function of $N_{\downarrow}/N_{\uparrow}$. Results for the zero-range model are depicted by filled squares: black  for $N_\uparrow = 33$, red for $N_\uparrow =57$, and blue for filled shells. Empty squares are the results for the square-well potential for the ranges: $R_0k_F=0.03({\color{green}\square});~0.12({\color{cpurple}\circ});~0.30({\color{red}\triangle});~0.60({\color{blue}\triangledown})$. The lines are the fits of \eq{eos} and the black dashed line is the zero-range equation of state with $F=0$. Inset: the coupling between the polarons in the Landau-Pomeranchuk quasiparticles model as a function of the range for zero-range model ($\blacksquare$) and square-well potential results ($\bullet$). Monte Carlo result of Pilati and Giorgini ($\color{blue}\triangledown$)~\cite{pil08} and the analytical result of Mora and Chevy ($\color{red}\blacktriangle$)~\cite{mora10}. The empty purple symbols represent $F=A^2/5$ and the green ones are values obtained by explicitly considering the energy differences between systems with 1 and 2 impurities (see text).}
\label{Fig:eosFIG}
\end{figure}

Energies of the Fermi gas as a function of impurity
concentrations $x=N_{\downarrow}/N_{\uparrow}$ are presented in
\fig{eosFIG}. The extrapolated value obtained with a square-well
potential range of $R_0 k_F = 0.03$ is in excellent agreement with the
one obtained by the zero-range model. Results in the range $ 0 \le x <
0.5$ were fitted to \eq{eos} considering previously estimated values of
$A$ and $m^*/m$. The dashed line in \fig{eosFIG} depicts the
curve of \eq{eos} with the term $x^2$ neglected ($F=0$) for the
zero-range model. A comparison of this curve and the fitted one shows
that as the concentration $x$ increases
to values above approximately $0.2$, effects of the Pauli blocking due to
the formation of a Fermi sea of the minority down-spin particles start
to add to the polaron density mediated interactions.

Simulations are performed for different concentrations of impurities keeping size effects under control.  In particular we have checked the equation of state for a Fermi sea of $N_\uparrow =33$ and $N_\uparrow =57$ atoms.
These results follow the trend shown by simulations made with
$N_\uparrow = 33$ particles, as we can see in \fig{eosFIG}. It is also
interesting to mention that there is no appreciable impact in the
results regarding whether or not the number of minority atoms corresponds to a closed
shell.

The strength of the interaction between the quasiparticles is accounted for
 the parameter $F$, which is obtained by fitting the energy for different impurity concentrations [see \eq{eos}]. Values of $F$ are estimated both for the zero-range model and at different ranges of the potential, as displayed in the inset of \fig{eosFIG}. Alternatively, the parameter $F$ can
also be estimated by subtracting the energy of the system with $M$ impurities and the energies of the systems with each single impurity immersed in the bath. In particular, for two impurities, we have
\begin{eqnarray}
 Fx^2&=& E(N_{\downarrow}=2;k=\{ 0,2\pi/L\})\\
 &-& \left( E(N_{\downarrow}=1;k= 0) + E(N_{\downarrow}=1;k=2\pi/L)\right) . \nonumber 
\end{eqnarray}
 These results are also shown for the zero-range model and for
the square-well potential as a function of the range for the concentration of impurities $x=2/33$. Both estimations of $F$
agree with each other. 

The result obtained via the zero-range model $F=0.22(8)$ is in
agreement with the one by Pilati and Giorgini~\cite{pil08}, $F = 0.14$, which also uses  quantum  Monte Carlo (QMC) calculations for closed-shell cases and fit the equation of state in the range of impurity concentration from 0 to 1. A better nominal agreement, $F=0.20$, occurs when a diagrammatic method is employed, as reported by Mora and Chevy~\cite{mora10}. A relation for the dominant polaron interactions in terms of a single polaron parameter in terms of the universal parameters $A$ was investigated in Ref.~\onlinecite{mora10}. For the normal state of the Fermi gas, this relation reads $F=A^2/5$. In the inset of \fig{eosFIG}, values of $A$ are those we computed for the zero-range model and for the square-well potential as a function of its range.

All data results obtained in this work are explicitly given in the Supplemental Material~\cite{pes21sup}.

\section{Conclusions}

Properties of a unitary Fermi polaron were studied using the
zero-range model where a contact potential is replaced by the
Wigner-Bethe-Peierls boundary condition in the $N$-body wave function. In addition these properties were estimated using a square-well potential as a function of its range. 

Simulations were performed to study finite-size effects and the results show that our results are free from these effects. Estimations of properties such as the polaron binding energy, effective mass, contact, and the role of different concentrations were found to depend on the range of the potential. Consistency in the results was exhaustively tested in particular by comparing estimations made with the zero-range model with those of a square-well potential as a function of its range. Although it is possible to argue that results obtained with a short-range potential using a small enough range give results within statistical uncertainties that might be equivalent to those obtained using the zero-range model, our estimates show that calculations made with a finite range are not as accurate as they could be. Moreover, difficulties in the simulations using very small ranges in effective potentials make it more convenient to use the zero-range model in simulations of ultracold gases in a unitary regime 

In summary, we have observed that using an effective potential with finite range can impact the estimation of the  ground-state properties of Fermi polarons. Nevertheless, estimated properties obtained by extrapolating the range of the square-well potential to the zero-range limit show agreement with the zero range  model. Our work shows that finite-range effects may be sizable, for instance, in a polaron with long-range interaction such as dipolar, Rydberg, or ionic polarons~\cite{Kleinbach18,Camargo18,Astrakharchik21}.\\

\section{Acknowledgments}

We thank Dr. Airton Deppman for his support of our work. Computations were performed at the Laboratório de Computação Científica - Universidade Federal de Goiás, at facilities provided by the project INCT-FNA Proc. No. 464 898/2014-5 and at the Centro Nacional de Processamento de Alto Desempenho em São Paulo (CENAPAD-SP). S.V. acknowledges financial support from the Brazilian agency, Fundação de Amparo \`{a} Pesquisa do Estado de São Paulo (FAPESP), project Proc. No. 2016/17612-7.

\appendix
\section{
Variational and diffusion Monte-Carlo method.}

The variational Monte Carlo (VMC) estimates the variational energy by sampling configurations from the probability density associated to the model wave function, given by \eq{psi}, using the Metropolis \textit{et al.}~\cite{met53} algorithm. The best set of variational parameters is obtained by minimizing the variational energy. The random displacement of the particles is adjusted so that the acceptance of new configurations is approximately $50\%$.

Configurations drawn in a VMC calculation are used in a diffusion Monte
Carlo (DMC) calculation; the Schr\"odinger's equation in imaginary time
$\tau=it/\hbar$,
\begin{equation}
    -\frac{\partial \psi(R;\tau)}{\partial \tau}= \left(H - E_0 \right)\psi(R;\tau),
    \label{eq:diff_schr}
\end{equation}
is a diffusion equation. A guess of the true energy $E_0$ is inserted in
\eq{diff_schr} to control the wave-function norm. In practice, the
variational energy can be a good starting choice for $E_0$, which is
periodically updated. Typically, \eq{diff_schr} is solved for the
lowest-energy state compatible with the given nodal structure of
\eq{psi} by defining a propagator at a small time step $\Delta\tau$ such
that the evolution in imaginary time is accomplished by successive
applications of this propagator~\cite{rey82,cha04}. The way to employ
this idea is to write a propagation equation,
\begin{widetext}
\begin{eqnarray}
    \Psi(R) \psi(R;\tau+\Delta\tau) =\int dR'\frac{\Psi(R)}{\Psi(R')}G(R,R';\Delta\tau) \Psi(R')\psi(R';\tau) 
    \label{eq:propag}
\end{eqnarray}
\end{widetext}
where the model wave function is used as a guide function and
$G(R,R';\Delta \tau)$ is a Green's function.
It is possible to show that the lowest-energy state can be projected out from the model wave function asymptotically in the imaginary time $\tau$~\cite{fou01,cha04} under the assumptions made using the above methodology. Within the limit of large imaginary time, the energy of the system is calculated using the mixed estimator defined as 
\begin{equation}
    E=\frac{\langle \Psi \vert H \vert \psi \rangle}{\langle \Psi
\vert\psi \rangle}.
\end{equation}
The time step in \eq{propag} must be small enough to avoid bias in the calculation of the energy and other properties, typically  $\Delta \tau \leqslant
10^{-4}~\frac{5}{3}E_{F\uparrow}^{-1}$~\cite{pes19}. The computational cells are constructed with $N_{\uparrow}= 33$ and $57$ up-spin atoms in the bath. Periodic boundary conditions are enforced in the simulations. Typically, the DMC calculations are conducted so that the system is evolved until $\tau \frac{3}{5}E_{F\uparrow}=3.0$ before starting to accumulate quantities of interest.
Quantities that do not commute with the Hamiltonian are estimated
through the mixed estimators, $Q_{\rm mix} = 2Q_{\rm DMC} - Q_{\rm VMC}$.

As discussed in a previous work~\cite{pes15} for the unpolarized Fermi
gas, the use of the zero-range boundary condition in the wave function
can insert divergences in the energy calculations. To solve the
divergences problems, we perform an additional move when sampling the particles' configurations. Essentially, after the random movement
of all the atoms in the gas, we promote the positions exchange of the
closest pair of particles with different spins. The configuration with
the exchanged pair is considered in the calculation with its respective probability.

\bibliographystyle{apsrev4-2}

\end{document}